\documentclass[aps,showpacs,showkeys,secnumarabic,12pt]{revtex4}
\usepackage{amsmath,amsbsy}

\newcommand{\pni}{\par\noindent}
\begin{document}
\title{On thin-shell wormholes evolving in  flat FRW 
spacetimes} \author {M. La Camera} \email{lacamera@ge.infn.it} 
\affiliation {Department of Physics and INFN - University of 
Genoa\\Via Dodecaneso 33, 16146 Genoa, Italy}
\begin{abstract} \pni
We analize the stability of a class of thin-shell wormholes with 
spherical symmetry evolving in flat FRW spacetimes.  The 
wormholes considered here are supported at the throat by a 
perfect fluid with equation of state $\mathcal{P}=w\sigma$ and 
have a physical radius equal to $aR$, where $a$ is a 
time-dependent function describing the dynamics of the throat and
$R$ is the background scale factor. The study of wormhole 
stability is done by  means of the stability analysis of 
dynamic systems. 
\end{abstract} 
\maketitle  
\section{I\lowercase{ntroduction}} \pni
Wormhole physics has become a popular research topic since the 
classical paper by Morris and Thorne.${}^{1}$ Early work was 
reviewed in the book of Visser ${}^{2}$ and there is an 
extensive recent review by Lobo.${}^{3}$ The class of thin-shell 
wormholes provides a wide collection of examples for the study 
of traversable Lorentzian wormholes. Visser ${}^{4,5}$ studied 
wormholes in Schwarzschild and Reissner-Nordstr$\ddot{o}$m 
backgrounds using for the first time the cut-and-past technique. 
The stability of static and dynamic wormholes  has been examined 
either by choosing specific equations of state or by considering 
a linearized stability analysis around a static solution. The 
literature is extremely extensive but for some examples see Refs.
6-19 and references therein. In this work we analyze the 
stability of a class of thin-shell wormholes with spherical 
symmetry evolving in a flat Friedmann-Robertson-Walker (FRW) 
cosmological background. Dynamic Lorentzian wormholes connecting 
FRW spacetimes were first studied by Hochberg and Kephart 
${}^{7}$ who discussed a possible resolution of the horizon 
problem using a network of evolving wormholes present in the 
early universe. We utilize the wormhole equations of motion given
in Ref. 7 and assume that the  matter at the shell is a 
perfect fluid with equation of state $\mathcal{P}=w\sigma$. We 
shall obtain an evolution equation for the physical radius $aR$ 
of the wormhole, where $a$ is a time-dependent function 
describing the dynamics of the throat and $R$ is the background 
scale factor. The study of stability is done considering the 
wormhole physical radii as dynamic systems and analyzing the 
behavior of their steady states with respect to a set of control 
parameters affecting the evolution.
\section{T\lowercase{hin-shell 
wormholes in cosmology}}\pni The dynamic wormholes we shall 
consider here result from surgically grafting two FRW spacetimes 
as proposed by Hochberg and Kephart ${}^{7}$ following the method
used by Visser ${}^{4,5}$ for constructing wormholes by cutting 
and pasting two manifolds to form a geodesically complete new 
manifold with a shell placed in the joining surface. The 
construction of Ref. 7, which we now briefly recall, starts 
taking two copies of the FRW solution 
\begin{equation} 
dS^2 = R^2(t)\,\left(\dfrac{dr^2}{1-k\,r^2}+r^2\,(d\vartheta^2+  
\sin^2 \vartheta d\varphi^2)\right)-dt^2 
\end{equation}
and then removes from every one of them an identical 
four-dimensional region of the form $\Sigma_\pm=\{r_\pm\leq 
a\}$ where $a=a(t)$ is a time-dependent radius. The resulting 
manifold contains two disjoint boundaries $\Sigma_\pm=\{r_\pm=a\}$
and by the identification $\Sigma_+ =\Sigma_{-} \equiv\Sigma$ 
one obtains two FRW spacetimes connected by a traversable 
wormhole with spherical symmetry whose throat is located on their
mutual boundary $\Sigma=\{f(r,t)=r-a(t)=0\}$. The corresponding 
geometry can be analyzed using the Israel-Lanczos-Sen 
${}^{20-22}$ ``thin-shell'' formalism of General Relativity. The 
wormhole throat is a timelike hypersurface with interior 
coordinates $\xi^i = (\vartheta,\varphi, \tau)$, $\tau$ being 
the throat proper time. The position of the throat in the 
background FRW spacetime is given by $X^\mu = (a(t),\vartheta, 
\varphi,t)$. The line element (1) with the curvature term $k$  
neglected is well justified to describe not only, as in Ref. 7,  
the  early stages of expansion but also the universe today on 
large scales,   so we shall put  $k=0$ in it. Moreover another 
realization of an evolving  wormhole was obtained by imagining 
its asymptotically parts as constituting flat FRW spacetimes. 
${}^{9}$  The tangent vectors to $\Sigma$ have the components 
$e^\mu_{(i)}\left|_\pm \right. = \partial X^\mu_\pm/\partial 
\xi^i$ and the induced metric $g_{ij} = e^\mu_{(i)}\,e^\nu_{(j)} 
\,g_{\mu\nu}\left|_\pm \right.$ on the junction hypersurface 
gives the line element 
\begin{equation}
ds^2 = a^2 R^2\,(d\vartheta^2+\sin^2 \vartheta d\varphi^2)- 
d\tau^2 
\end{equation}
having set 
$\left[-\left(\dfrac{dt}{d\tau}\right)^2+\left(\dfrac{d}{d\tau}
a(t)\right)^2R^2(t)\right]=-1$ the coefficient of $d\tau^2$. 
Notice that the physical radius of the wormhole is $aR$.  Using 
$a' =\dot{a}\,\dfrac{dt}{d\tau}$, where the prime and 
the dot represent the derivatives with respect to $\tau$ and 
$t$ respectively,  one obtains
\begin{equation} 
\dfrac{dt}{d\tau}=\dfrac{1}{\sqrt{1-(\dot{a}R)^2}}
\end{equation}
therefore it will be possible in the following to eliminate one 
of the two time parameters $\tau$ and $t$ in favour of the other.
The second fundamental form (extrinsic curvature) is given by
\begin{equation}    
K^\pm_{ij}=-n_\gamma\,\left(\frac{\partial^2 X^\gamma}{\partial 
\xi^i \partial \xi^j}+\Gamma^{\gamma\pm}_{\alpha\beta}\,
\frac{\partial X^\alpha}{\partial \xi^i}\frac{\partial X^\beta}
{\partial \xi^j}\right) 
\end{equation} 
where $n_\gamma$ is the unit normal to $\Sigma$: 
\begin{equation}
n_\gamma = \pm\, \left|g^{\alpha\beta}\frac{\partial f}{\partial 
X^\alpha}\frac{\partial f}{\partial X^\beta}\right|^{-1/2}\, 
\frac{\partial f}{\partial X^\gamma} \quad .
\end{equation}
With the definitions $\kappa_{ij}=K^+_{ij}-K^{-}_{ij}$ the 
Lanczos equations follow from the Einstein field equations for 
the hypersurface, and are given by
\begin{equation}
S^i_j = -\,\frac{1}{8\pi}\,\left(\kappa^i_j -\, 
\delta^i_j\,\kappa^h_h \right)
\end{equation}
where $S^i_j$ is the surface stress-energy tensor on $\Sigma$. In
the case of spherical symmetry considerable simplifications 
occur, namely $\kappa_{ij}\,=\, \textrm{diag}\left(\kappa^\vartheta_ 
\vartheta,\,\kappa^\vartheta_\vartheta,\,\kappa^\tau_\tau\right)$,
and the surface stress-energy tensor may be written in terms of 
the surface pressure $\mathcal{P}$ and the surface energy density
$\sigma$ as $S^i_j \,=\,\textrm{diag}\left(\mathcal{P},\mathcal{P},
-\,\sigma\right)$. The Lanczos equations then reduce to
\begin{eqnarray}
\mathcal{P}&=& \frac{1}{8\pi}\,\left(\kappa^\vartheta_\vartheta\,+ 
\,\kappa^\tau_\tau\right)\\ 
\sigma&=&-\,\frac{1}{4\pi}\,\kappa^\vartheta_\vartheta 
\end{eqnarray}
where
\begin{eqnarray}
\kappa^\vartheta_\vartheta &=& a'\dot{R}+\dfrac{\sqrt{1+(a'R)^2}}
{aR} \\
\kappa^\tau_\tau &=& \dfrac{a''R}{\sqrt{1+(a'R)^2}} +2a'\dot{R}
\quad. 
\end{eqnarray}
Using  Eq. (3), it is possible to  substitute   derivatives 
with respect to the proper time on the shell $\tau$ with 
derivatives with respect to the cosmic time $t$, so the 
components of the tensor $S^i_j$ are writtem as 
\begin{eqnarray}
\mathcal{P} &=& \dfrac{\left[1-(\dot{a}R)^2\right]^{-1}\,\dfrac{
d}{dt}\,(\dot{a}R)+2\dot{a}\dot{R}+\dfrac{1}{aR}}{4\pi\,\sqrt{1-
(\dot{a}R)^2}} \\
\sigma &=& -\,\dfrac{\dot{a}\dot{R}+\dfrac{1}{aR}}{2\pi\,\sqrt{1-
(\dot{a}R)^2}} 
\end{eqnarray} 
and the conservation identity $S^i_{j;i}=0$, which is not 
independent but can be obtained from the field equations,  
provides the simple relationship
\begin{equation}
\dot{\sigma}+2\,\left(\dfrac{\dot{a}}{a}+\dfrac{\dot{R}}{R}\right)
\,(\mathcal{P}+\sigma)=0.
\end{equation}
If one chooses a particular equation of state, in the form
$\mathcal{P}=\mathcal{P}(\sigma)$, then the conservation equation 
can be formally integrated obtaining
\begin{equation}
\displaystyle{\int\dfrac{d\sigma}{\mathcal{P}(\sigma)+\sigma}}+2\,
\ln{(aR)}=0 \quad.
\end{equation}
The choice $\mathcal{P}=\mathcal{P}(\sigma)$, because of Eqs. (11)
and (12), will set a bond between $a$ and $R$. In this paper we
shall consider the case 
\begin{equation}
\mathcal{P}= w \sigma
\end{equation}
where $w$ is an arbitrary real constant. Therefore the  
conservation equation (14) and the equation of state (15) become 
respectively 
\begin{eqnarray} 
&{}&\sigma =  \sigma_0\,\left(\dfrac{a_0R_0}{aR}\right)^{2(1+w)}\\ 
&{}&\left[1-\,(\dot{a}R)^2\right]^{-1}\,\dfrac{d}{dt}(\dot{a}R)+ 
2\,(1+w)\,\dot{a}\dot{R}+\dfrac{(1+2w)}{aR} = 0 
\end{eqnarray} 
where the subscript ${}_0$ refers to quantities calculated at 
some initial time $t=t_0$. We have four unknowns: $a,\, R,\, 
\mathcal{P}$ and $\sigma$, but only  three independent equations:
(11), (12) and (15). The standard stability  method based on the 
definition of a potential ${}^{8}$ is not applicable here because
from those equations it does not appear possible to obtain 
explicitly the derivative with respect to the time of the 
physical wormhole radius $aR$ as a function of $aR$. Thus, to 
close the system, we need to have one more equation. A suitable 
fourth equation must be chosen in such a way to facilitate the 
stability analysis  for the examined class of wormholes. 
\section{S\lowercase{tability analysis}}\pni 
In this section, to analyze the stability of the wormhole 
physical radius we substitute again, using Eq. (3), derivatives
of $aR$ with respect to $\tau$ with derivatives with respect to 
$t$ and it may be checked that at equilibrium,  where 
$\dfrac{d}{d\tau}(aR)$ and $\dfrac{d}{dt}(aR)$ vanish,  the 
final result with regard to the stability does not change. We 
consider the wormhole radius $aR$ as a dynamical system (see, 
e.g., Ref. 23) with an evolution equation of the form 
\begin{equation} 
\dfrac{d}{dt}\,(aR) = F(aR, \lambda) 
\end{equation} 
where $F$ is  a differentiable function acting on the spacetime 
where $aR$ is defined and $\lambda$ denotes a set of control 
parameters affecting the evolution. The value of $aR$ for which 
$F(aR, \lambda)=0$ is a steady state point. The stability of the 
equilibrium will be studied considering small perturbations away 
from the above fixed point and determining, by means of a 
linear analysis,  the tendency of the perturbations to grow or to
decay in time. Now 
\begin{equation} 
\dfrac{d}{dt}\,(aR) = \dot{a}R+a\dot{R} 
\end{equation}
so we must know both  $\dot{a}R$ and $a\dot{R}$  as a 
function of $aR$. Here it is worth noticing that when 
$w=-1/2$ the equation of state (17) becomes
\begin{equation}
\left[1-\,(\dot{a}R)^2\right]^{-1}\,\dfrac{d}{dt}(\dot{a}R)+
\dot{a}\dot{R}=0
\end{equation}
which is satisfied either when the throat function $a$ is 
constant, and so $\dot{a}=0$, or when $a$ is time-dependent, 
and $\dot{a}$ is given integrating Eq. (20):
\begin{equation}
|\dot{a}| = \dfrac{1}{R}\left\{1+\dfrac{[1- (\dot{a}_0R_0)^2]} 
{(\dot{a}_0R_0)^2}\left(\dfrac{R}{R_0} \right)^{2}
\right\}^{-1/2}. 
\end{equation}
When $\dot{a}=0$, the wormhole  radius $aR$ simply varies  in 
direct proportion to the scale factor $R$. When $\dot{a}\neq 0$,
which is the case treated in this paper,  the behavior  of $aR$  
can be studied  if the scale factor $R$ is known. When  $w\neq 
-1/2$, to choose  a suitable fourth equation  we 
consider a specific class of simple wormhole solutions 
corresponding to the choice 
\begin{equation} 
\left[1-\,(\dot{a}R)^2\right]^{-1}\,\dfrac{d}{dt}(\dot{a}R) -
2w\, \dot{a}\dot{R} =0 
\end{equation}
which generalizes Eq. (20) to the other values of $w$, and the 
equation of state (17) becomes 
\begin{equation}
(1+2w)\,\left(2\,\dot{a}\dot{R}+\dfrac{1}{aR}\right)=0 \quad.
\end{equation}
In the following we shall consider values  of $w$ different from 
the value $w = -\,1/2$ which was briefly discussed above, so Eq. 
(23) gives  
\begin{equation}
2\,\dot{a}\dot{R}+\dfrac{1}{aR}=0 \quad.
\end{equation}
Then we shall treat a particular  class of wormholes for which  
$\dot{a}$ and $\dot{R}$ must have opposite signs and therefore 
when one is increasing the other is decreasing. Now we 
substitute $a\dot{R}=-1/(2\,\dot{a}R)$ into Eq. (19) which 
becomes 
\begin{equation} 
\dfrac{d}{dt}\,(aR) = \dot{a}R - \dfrac{1}{2\,\dot{a}R} 
\end{equation} 
therefore the equilibrium is  reached when $(\dot{a}R)^2=1/2$. 
Moreover the components of the tensor $S^i_j$ now become 
\begin{eqnarray} 
\mathcal{P}&=&-\,\dfrac{w}{4 \pi a R 
\sqrt{1-(\dot{a}R)^2}}\\ \sigma &=&-\,\dfrac{1}{4 \pi a R 
\sqrt{1-(\dot{a}R)^2}} 
\end{eqnarray} 
so the weak energy  condition is violated, while the null energy 
condition is satisfied when $w \leq -\,1$. Finally, to obtain 
$\dot{a}R$ as a function of $aR$ we equate Eqs. (16) and (27) for
the surface density $\sigma$ and have 
\begin{equation} 
1+4\pi a_0 R_0\sigma_0 \sqrt{1-\ 
(\dot{a}R)^2}\,\left(\dfrac{a_0R_0}{aR}\right)^{1+2w}=0 \quad. 
\end{equation}
Evaluating the previous equation at the time $t=t_0$ it results 
\begin{equation}
1+4 \pi a_0 R_0\sigma_0 \sqrt{1-\ (\dot{a_0}R_0)^2}=0 \quad.
\end{equation}    
Then Eq. (28) takes the form
\begin{equation}
\sqrt{\dfrac{1-\,(\dot{a}R)^2}{1-\,(\dot{a_0}R_0)^2}} = 
\left(\dfrac{aR}{a_0R_0}\right)^{1+2w} 
\end{equation}
and therefore
\begin{equation} 
|\dot{a}R| =  \sqrt{1-\, [1-(\dot{a}_0R_0)^2]\left(\dfrac{aR}
{a_0R_0}\right)^{2(1+2w)}} \quad.
\end{equation}
Then, when $w=0$ both $\dot{a}R$, by Eq. (22), and $aR$ are
constant. Now we can write Eq. (25) in the form of Eq. (18):
\begin{equation}
\dfrac{d}{dt}\,(aR)= \textrm{sign}[\dot{a}]\,\dfrac{\dfrac{1}{2}
-\,[1-(\dot{a}_0R_0)^2]\left(\dfrac{aR}{a_0R_0}\right)^{2(1+2w)}}
{\sqrt{1-\,[1-(\dot{a}_0R_0)^2]\left(\dfrac{aR}{a_0R_0}
\right)^{2(1+2w)}}} 
\end{equation} 
where $\textrm{sign}[\dot{a}] = \dot{a}/|\dot{a}|$. 
The right-hand side vanishes when
\begin{equation}
a_*R_*=a_0R_0\left[2\,[1-(\dot{a}_0R_0)^2]\right]^{-1/(2(1+2w))}
\end{equation}
so $a_*R_*$ is the static solution which gives the radius of the 
throat at the equilibrium. If $(\dot{a}_0R_0)^2=1/2$,  then the 
static solution is $a_0R_0$ and does not depend on $w$. The role 
of  perturbation of stability will be expressed by setting in Eq.
(32) 
\begin{equation} 
aR=a_*R_* +\xi(t) 
\end{equation} 
where  $|\xi|\ll a_*R_*$ represents the disturbance. We are 
interested in linear analysis so higher order terms will be 
neglected in the Taylor expansion of the evolution equation (32) 
which yields at first order 
\begin{equation} 
\dfrac{d}{dt}\,\xi=-\,\textrm{sign}[\dot{a}]\,\sqrt{2}\,(1+2w)\,
\dfrac{\xi}{a_*R_*}
\end{equation} 
Therefore we can conclude, at least until the nonlinear region is
reached, that :\pni (i) if  $(1+2w)\,\textrm{sign}\,[\dot{a}]>0$,
then the perturbation $\xi(t)$ decays exponentially and the 
equilibrium is stable; (ii) if  $(1+2w)\,\textrm{sign}\,
[\dot{a}]<0$, then the perturbation  $\xi(t)$ grows exponentially
and the equilibrium is unstable. Here $\dot{a}$ and
$\dot{R}$ have opposite signs, so in an expanding universe the 
equilibrium is stable when $w<-1/2$. Finally we recall that,  
having chosen the equation of state in the form
$\mathcal{P}=\mathcal{P}(\sigma)$, the functions $a$ and $R$ are 
related. This can be seen explicitly integrating first Eq. (22): 
\begin{equation}
|\dot{a}R| = \left\{1+\dfrac{[1- 
(\dot{a}_0R_0)^2]}{(\dot{a}_0R_0)^2}\left(\dfrac{R}{R_0} 
\right)^{-4w}\right\}^{-1/2} 
\end{equation}
and then equating, in the case $w\neq -1/2$, Eqs. (31) and (36).
The result is 
\begin{equation}
\dfrac{a}{a_0} =\dfrac{R_0}{R}\, \left\{[1-(\dot{a_0}R_0)^2] 
+(\dot{a}_0R_0)^2 \left(\dfrac{R}{R_0}\right)^{4w}\right\}^ 
{-1/(2(1+2w)} 
\end{equation}
\section{C\lowercase{onclusion}}\pni In this paper we have 
considered a particular  class of thin-shell wormholes evolving 
in a flat FRW cosmological background and analized their 
stability to linearized perturbations  around static solutions 
using the  stability analysis of dynamical systems. The physical 
radius of the wormholes is equal to $aR$, where $a$ is a 
time-dependent function describing the dynamics of the throat 
and $R$ is the background scale factor. We have considered the 
equation of state $\mathcal{P}=w\sigma$ and found that the factor
$w$   is crucial for  determining the stability of the 
equilibrium.   Wormholes supported by other equations of state 
than the one used in this paper can be used to generate other 
families of additional solutions. 
    

\begin{thebibliography}{00} 
\bibitem{1} M. S. Morris and K. S. Thorne, \textit{Am. J. Phys.} 
\textbf{56}, 395 (1988). 
\bibitem{2} M. Visser,  \textit{Lorentzian wormholes from 
Einstein to Hawking} (AIP Press, New York, 1995). 
\bibitem{3} F. S. N. Lobo, \textrm{in} \textit{Classical and  
Quantum Gravity Research Progress}, 1 (Nova Science Publishers, 
2008).
\bibitem{4} M. Visser, \textit{Phys. Rev. D} \textbf{39}, 3182  
(1989).  
\bibitem{5} M. Visser \textit{Nucl. Phys. B} \textbf{328}, 203
(1989).
\bibitem{6} M. Visser, \textit{Phys. Lett. B} \textbf{242}, 24
(1990). 
\bibitem{7} D. Hochberg and T. W. Kephart, \textit{Phys. Rev. 
Lett.} \textbf{20}, 2665 (1993). 
\bibitem{8} E. Poisson and M. Visser, \textit{Phys. Rev. D} 
\textbf{52}, 7318 (1995).
\bibitem{9} S. Kar and D. Sahdev,  \textit{Phys. Rev. D} 
\textbf{53}, 722 (1996).
\bibitem{10} E. F. Eiroa and G. E. Romero, \textit{Gen. Rel. 
Grav.} \textbf{36}, 651 (2004). 
\bibitem{11} F. S. N. Lobo and P. Crawford, \textit{Gen. Rel. 
Grav.} \textbf{21}, 391 (2004). 
\bibitem{12} F. S. N. Lobo, \textit{Class.Quant. Grav.} 
\textbf{21},  4811 (2004). 
\bibitem{13} K. A. Bronnikov and S. V. Grinyok, \textit{Grav. 
Cosmol.} \textbf{11}, 75 (2005).
\bibitem{14} E. F. Eiroa and C. Simeone, \textit{Phys. Rev. D} 
\textbf{76}, 024021 (2007).
\bibitem{15} J. P. S. Lemos and F. S. N. Lobo, \textit{Phys. Rev.
D} \textbf{78}, 044030 (2008).
\bibitem{16} F. Rahaman, M. Kalam and S. Chakraborty, 
\textit{Gen. Rel. Grav.} \textbf{38}, 1687 (2008).
\bibitem{17} E. P. Eiroa, \textit{Phys. Rev. D}  \textbf{78}, 
024018 (2008).  
\bibitem{18} P. K. F. Kuhfittig, \textit{Acta Phys. Polonica B} 
\textbf{41}, 2017 (2010).
\bibitem{19} G. A. S. Dias and J. P. S. Lemos, [arXiv:1008.3376 
[gr-qc]].  
\bibitem{20} N. Sen, \textit{Ann. Phys.} (Leipzig) \textbf{73}, 
365 (1924).
\bibitem{21} K. Lanczos, \textit{Ann. Phys.} (Leipzig) 
\textbf{74}, 518 (1924). 
\bibitem{22} W. Israel, \textit{Nuovo Cimento} \textbf{44 B}, 1
(1966); \textbf{48 B}, 463(E) (1967).    
\bibitem{23} C. G. B$\ddot{o}$mer, T. Harko and S. V. Sabau,
[arXiv:1010.5464 [gr-qc]].  
\end{thebibliography}
\end{document}